\documentclass[%
 reprint,
superscriptaddress,
 amsmath,amssymb,
 aps,
]{revtex4-2}

\usepackage{graphicx}%
\usepackage{dcolumn}%
\usepackage{bm}%

\usepackage{amsmath}
\usepackage{graphicx}
\usepackage[colorlinks=true, allcolors=blue]{hyperref}
\usepackage{ulem}

\usepackage{amsthm}
\newtheorem{theorem}{Result}

\usepackage{algorithm}
\usepackage{algpseudocode}

\usepackage{physics}
\usepackage{xcolor}

\DeclareMathOperator*{\argmin}{arg\,min}

\newcommand{\half}{\frac{1}{2}}
\newcommand{\roothalf}{\frac{1}{\sqrt{2}}}
\newcommand{\Var}[1]{\text{Var}[#1]}
\newcommand{\veps}{\varepsilon}
\newcommand{\blue}[1]{{\color{blue}{#1}}}
\newcommand{\gray}[1]{{\color{gray}{#1}}}

\newcommand{\m}{\mathcal}

\newcommand{\ep}[1]{^{(#1)}}

\newcommand{\knoclid}{\text{$k$-NoCliD}}
\newcommand{\noclid}{\text{NoCliD}}

\begin{document}

\title{Non-Clifford diagonalization for measurement shot reduction in quantum expectation value estimation}%

\author{Nicolas PD Sawaya}
\email{nicolas@azulenelabs.com}
\affiliation{\azulene}

\author{Daan Camps}
\affiliation{\nersc}

\author{Ben DalFavero}
\affiliation{\msu}

\author{Norm M Tubman}
\affiliation{\nasa}

\author{Grant M Rotskoff}
\affiliation{\stanford}

\author{Ryan LaRose}
\email{rmlarose@msu.edu}
\affiliation{\msu}

\newcommand{\azulene}{Azulene Labs, San Francisco, CA 94115}
\newcommand{\msu}{Center for Quantum Computing, Science, and Engineering, Michigan State University, East Lansing, MI 48823, USA}
\newcommand{\stanford}{Department of Chemistry, Stanford University, Stanford, California 94305, USA}
\newcommand{\nersc}{National Energy Research Scientific Computing Center, Lawrence Berkeley National Laboratory, Berkeley, California, USA}
\newcommand{\nasa}{NASA Ames Research Center, Moffett Field, CA, 94035, USA}

\date{\today}%

\begin{abstract}
Estimating expectation values on near-term quantum computers often requires a prohibitively large number of measurements. One widely-used strategy to mitigate this problem has been to partition an operator's Pauli terms into sets of mutually commuting operators. Here, we introduce a method that relaxes this constraint of commutativity, instead allowing for entirely arbitrary terms to be grouped together, save a locality constraint. The key idea is that we decompose the operator into arbitrary tensor products with bounded tensor size, ignoring Pauli commuting relations. This method --- named $k$-NoCliD ($k$-local non-Clifford diagonalization) --- allows one to measure in far fewer bases in most cases, often (though not always) at the cost of increasing the circuit depth. We introduce several partitioning algorithms tailored to different Hamiltonian classes. For electronic structure, we numerically demonstrate the existence of threshold values of $k$ for which \knoclid{} leads to the lowest shot counts, though we leave improved partitioning algorithms to future work. We focus primarily on three Hamiltonian classes---molecular vibrational structure, Fermi-Hubbard, and Bose-Hubbard---and show that $k$-NoCliD reduces the number of circuit shots, often by a very large margin, and often even for $k$ as small as 2.
\end{abstract}

\maketitle

\tableofcontents

\section{Introduction}

Calculating expectation values of observables is a ubiquitous subroutine in quantum computing~\cite{McArdle:2020:015003,Bauer:2020:12685,PRXQuantum.3.020323}. Importantly, for some application areas like chemistry and materials science~\cite{mat2,mat3,eriksen2020ground}, an enormous number of measurements (i.e. circuit repetitions or circuit ``shots'') are required to estimate expectation values to reasonable accuracy \cite{crawford2021si,gonthier2022meas,yen23deterministic,mat4}.

In qubit space, the Hamiltonian of interest can be represented as a linear combination of Pauli terms,
\begin{equation}\label{eq:generalham}
H = \sum_i c_i P_i,
\end{equation}
where $P_i \in \{I,X,Y,Z\}^{\otimes n}$, $n$ is the number of qubits, $\{I,X,Y,Z\}$ are the Pauli matrices, and $c_i$ are real coefficients.

Several algorithms have been developed for reducing the number of circuit shots required \cite{izmaylov2019anticomm, verteletskyi20qwc, chong20fc, zhao2020measurement, huang2020predicting, crawford2021si, yen21cartan, huggins2021efficient, choi2022ghost, choi2023fluid, yen23deterministic, majland2023optimizing, dutt2023practical,dalfavero2023kqwc}. Perhaps the most widely used class of algorithms for shot reduction is based on partitioning $H$ into sets of commuting Pauli terms \cite{chong20fc,bonet2020nearly,crawford2021si,yen23deterministic,inoue2024almost}. Both the commutation locality (qubit-wise commuting \cite{verteletskyi20qwc}, k-qubitwise \cite{dalfavero2023kqwc}, or fully commuting \cite{chong20fc}) and the partition algorithm (e.g. SortedInsertion \cite{crawford2021si} or graph-based methods \cite{verteletskyi20qwc}) have been studied. Such partitioning strategies have been shown to reduce the shot counts by orders of magnitude, though even after this reduction the required shots may be prohibitive \cite{gonthier2022meas}.

Here we introduce a novel partition-and-diagonalize strategy that does \textit{not} require the Pauli terms in each partition to commute. We introduce the $k$-NoCliD ($k$-local non-Clifford diagonalization) algorithm. The basic idea is to transform equation \eqref{eq:generalham} into a partition of product states of arbitrarily sized tensors, and measure in the basis of each individual tensor product. This allows one to measure in far fewer bases and leads to far fewer circuit shots in many cases, as we numerically demonstrate for a broad set of Hamiltonians.

\section{Theory}

A quantum operator may be expressed as 
\begin{equation}\label{eq:H_h_q}
H = \sum_{q = 1}^{K}  h_q,
\end{equation}
such that each operator $h_q$ may be measured in a single basis, with many circuit shots yielding an estimate for $\expval{h_q}$. At this point, equation \eqref{eq:H_h_q} makes no assumptions whatsoever about the form of $h_q$. 
(If the Hamiltonian is expressed as a weighted Pauli string as in equation \eqref{eq:generalham}, a commonly used strategy is to group Pauli strings $P_i$ that satisfy some mutual commutativity relation into the same operator $h_q$, as discussed below.)  %
The general expression %
for total measurements $N_{tot}$ required for error $\veps$  on $\expval{H}$ is \cite{crawford2021si}
\begin{equation}\label{eq:generalshotcounts}
\varepsilon^2 N_{tot} = \left( \sum_{q = 1}^{K} \sqrt{\text{Var} [h_q]} \right)^2.
\end{equation}
In this work we focus on expectation value methods where shot counts scale as $O(1/\varepsilon^2)$, which are appropriate for nearer-term NISQ algorithms. We do not consider $O(1/\varepsilon)$-scaling methods \cite{knill2007optimal,obrien2022efficient,steudtner2023fault}, which are appropriate for long-term error-corrected quantum computers as they require much longer circuit depths.

The most widely-used methods \cite{chong20fc,crawford2021si,yen23deterministic} for estimating $\expval{H}$ have required that each $h_q$ be a superposition of Pauli strings $P_i$ that mutually commute. In this work we remove this constraint, allowing for measurement in an arbitrary ``non-Pauli'' basis. One consequence is that the basis-change circuits are no longer Clifford circuits \cite{crawford2021si}; hence we call the method \knoclid{}, for $k$-local non-Clifford diagonalization.

To begin justifying this approach of measuring in a non-Pauli basis, it is instructive to consider a one-qubit example. We consider a real Hamiltonian and real state, \textit{i.e.} we constrain ourselves to the $XZ$ plane. Consider the Hamiltonian
\begin{equation}
H_\eta=\eta X + \sqrt{1-\eta^2} Z,
\end{equation}
a unit-norm one-qubit operator for which the eigenvectors lie along an angle $\arccos{\eta}$ relative to the $Z$ axis. Then, we compare the traditional approach of measuring in two bases ($X$ and $Z$ bases) before reconstructing $\expval{H_\eta}=\eta \expval{X} + \sqrt{1-\eta^2} \expval{Z}$, versus measuring ``directly'' in just one non-Pauli bases or ``rotated basis'' $P_{H_\eta} = H_\eta$ (note that $P_{H_\eta}$ has unit norm, $P_{H_\eta} = I$). The latter is done by performing a rotation around the $Y$ axis by an angle $\arccos{\eta}$ before measurement. %

Appendix \ref{apx:thm1q} provides additional pedagogical discussion, where we also prove the following result. For a given error, $N_{tot}^{GPB}$ is the number of shots required when using the global Pauli basis, \textit{i.e.} reconstructing $\expval{H}$ from $\expval{X}$ and $\expval{Z}$. $N^{RB}_{tot}$ is the number of shots when measuring in the single rotated basis $\expval{H_\eta}$.
\begin{theorem}\label{thm:1q}
Consider evaluating $\ev{H}{\psi}$ for a Hamiltonian $H_\eta=\eta X + \sqrt{1-\eta^2} Z$ with $\eta \in [0,1]$ and any real state $\ket{\psi}=\alpha \ket{0} + \sqrt{1-\alpha^2} \ket{1}$, where the rotated basis is $P_{H_\eta}$=$H_\eta$. For all $\eta$ and $\alpha$, $N_{tot}^{RB} \leq N_{tot}^{GPB}$. Hence, in terms of shot counts, it is always optimal to measure in this rotated non-Pauli basis.
\end{theorem}

This finding for the simple one-qubit case motivates this work, where we explore whether using arbitrary non-Pauli bases is beneficial in larger problems.

\subsection{k-local non-Clifford diagonalization}

The procedure of \knoclid{} begins with refactoring the Hamiltonian into $L$ operators as 
\begin{equation}\label{eq:sumMq}
H = \sum_{q=1}^{L} M_q,
\end{equation}
where the key point is that each $M_q$ may be measured in a single basis and one would usually like to have $L \ll K$. The structure of each $M_q$ is discussed below. Thus one measures in $L$ bases in order to reproduce $\expval{H}=\sum_q \expval{M_q}$. To evaluate each $\expval{M_q}$ on a quantum computer, it is necessary to diagonalize each $M_q$ operator as 
\begin{equation}
M_q = U_{q} D_{q} U_{q}^{\dagger}.
\end{equation}

Given many copies of the state of interest $\ket{\psi}$ --- prepared for example by variational quantum eigensolver (VQE) \cite{mcclean2016theory} and sometimes denoted $\ket{\psi} = U_{VQE} \ket{0}$ --- to calculate $\expval{M_q}$ one must execute a quantum circuit for $U_q^{\dagger}$. This procedure leads to
\begin{equation}\label{eq:exped_UDU}
\begin{split}
\expval{M_q} &= \bra{\psi} U_{q} U_{q}^{\dag} M_{q} U_{q} U_{q}^{\dag} \ket{\psi} \\
&= \bra{\psi} U_{q} D_{q} U_{q}^{\dag}\ket{\psi}.
\end{split}
\end{equation}
Hence after implementing the change-of-basis circuit $U_q^{\dag}$, the calculation of $\expval{M_q}$ has been transformed to a calculation of $\expval{D_q}$, which can now be performed in the computational $Z$ basis because $D_q$ is diagonal. Note that in a full implementation one must consider the cost to compile the circuit unitary $U_{q}$ on a classical computer. The estimator for the Hamiltonian is then 
\begin{equation}
\expval{H} = \sum_q \expval{M_q} = \sum_q \bra{\psi} U_{q} D_{q} U_{q}^{\dag} \ket{\psi}.
\end{equation}

Previously proposed shot reduction methods \cite{chong20fc,crawford2021si,yen23deterministic} may be expressed as equation \eqref{eq:sumMq} (though they are not always expressed as such). For example, in fully commuting (FC) partitioning, $M_q= \sum_{P_j \in \m R_q} c_j P_j$, where all $P_j$ in the set $\m R_q$ mutually commute  \cite{chong20fc,yen23deterministic}. In FC, the diagonalization circuit $U_q^{\dag}$ is always a Clifford circuit requiring at most depth $O(n^2/\log n)$ gates  \cite{aaronson04stabilizer}. By contrast, in \knoclid{}, $U_q^{\dag}$ is \textit{not} constrained to be a Clifford circuit, which can lead to large reductions in shot counts, as demonstrated in our numerical section. Because we lose the guarantee that the depth scales quadratically, shot count reduction often comes with an increase in circuit depth. Generally, we write the circuit depth is $O(f(k))$ for some function $f(k)$. While compiling arbitrary unitaries results in $f(k) \sim 2^k$ \cite{knill1995approximation,iten2016isometries}, this may often be an overly pessimistic estimate because the operators to diagonalize in problems of interest are often highly structured (for example strings of bosonic position operators $q_i \otimes q_j \otimes \cdots$).
Furthermore, often the circuit depth will not increase (or the increase will be manageable) because \knoclid{} is intended to be implemented with $k \ll n$. This will often lead to $O(f(k))$ being less than $O(n^2/log(n))$. Further, while $n$ increases with system size by definition, \knoclid{} will often lead to $k$ being naturally bounded by a constant, especially for bosonic and phononic Hamiltonians, but also for the greedy and blocking algorithms introduced below. %

We now discuss the specific structure of $M_q$ that is used in \knoclid{}. The $M_q$ are chosen such that they can be factorized as
\begin{equation}\label{eq:Mdef}
M_q = \sum_{r=1}^{l_q} W_{qr}
\end{equation}
where each $W_{qr}$ is a tensor product
\begin{equation}\label{eq:ttrain}
W_{qr} = A_{qr1}^{\m K_{qr1}} \otimes A_{qr2}^{\m K_{qr2}} \otimes A_{qr3}^{\m K_{qr3}} \otimes ...
\end{equation}
such that $\m K_{qrs}$ is the set of number of qubits over which the local tensor acts, and we require the cardinality of each set to be $|\m K_{qrs}| \equiv k_{qrs} \leq k$. The $k_{qrs}$ may be unequal but they must all be at most $k$, and for any tensor product $W_{qr}$ we have $\sum_s k_{qrs} = n$, the number of qubits. Note that equation \eqref{eq:ttrain} does not require qubits to be adjacent. 
All tensor products $W_{qr}$ in the same partition $q$ must commute with each other, $\forall (r,r'): [W_{qr},W_{qr'}] = 0$; this constraint is what allows $\expval{M_q}$ to be measured in a single basis. Note that $W_{qr}$ is not restricted to be a Pauli term, so this commutation constraint is not the same as the one used in traditional shot reduction \cite{chong20fc}. For example, for 2-\noclid, $W_{qr} = (X_1 X_2 + Y_1 + Z_1) \otimes (Y_3 Z_4)$ and $W_{qr'} = (X_1 X_2 + Y_1 + Z_1) \otimes (I_3 I_4)$ do obey this constraint, where subscripts denote qubit index.

\textit{The key benefit of the method is that one need measure in only $L$ bases} (one for each $M_q$), where $L$ can be made to be much smaller than the number of partitions (\textit{i.e.} fragments) required in traditional methods. This in turn can lead to large reductions in shot counts. Note that a smaller number of fragments does not automatically guarantee lower shot counts \cite{yen23deterministic}---numerical results, provided below, are required to confirm that any given partitioning indeed leads to lower shot counts.

Because for a given $q$ all $W_{qr}$ commute, they are simultaneously diagonalizable. This work's numerical implementation will assume that \textit{tensor-wise commutation} holds, such that in a tensor product each tensor can be diagonalized independently. (A counter-example would be $q_i q_j$ and $p_i p_j$ where $q$ and $p$ are respectively bosonic position and momentum operators---the terms commute but they require diagonalization across two tensors). This constraint makes the formalism easier but it is not required for \knoclid{}. Thus the diagonalization unitaries take the form
\begin{equation}
U_{q} = V_{q,1} \otimes V_{q,2} \otimes \cdots
\end{equation}
where all $V$ are unitary and tensor sizes match those in equation \eqref{eq:ttrain}. For a given $M_q$, one can then implement all $V$ in parallel. Each $V$ may be decomposed into a gate set via a variety of circuit compilation methods \cite{knill1995approximation,iten2016isometries,bqskit}. The choice of $k$ is arbitrary; a larger $k$ will usually lead to fewer circuit shots but more circuit depth. %

Note that the \knoclid{} construction has one more ``layer'' of complexity than commonly used methods like QWC and full-commuting partitioning: we partition the Hamiltonian into distinct $M_q$ operators, which are in turn partitioned into multiple $W_{qr}$, which in turn may be represented as sums of Pauli strings.

Formally stated, in \knoclid{} the optimal partitioning for a Hamiltonian $H$ is
\begin{equation}
\begin{split}
&\argmin_{ \{W_{qr}\} } \sum_{q=1}^L \sqrt{\text{Var} [M_q]} \\
= &\argmin_{ \{W_{qr}\} } \sum_{q=1}^L \sqrt{\text{Var} [\sum_{r} W_{qr}]}
\end{split}
\end{equation}
subject to the constraints
\begin{equation}
\begin{split}
\sum W_{qr} &= H
\\
\forall  k_{qri}: k_{qri} &\leq k
\\
\forall (q, r,r'): [W_{qr},W_{qr'}] &= 0.
\end{split}
\end{equation}

This optimization is over a continuous space, since multiple $W$ tensors may include some component of the same Pauli string $P_i$ with arbitrary coefficient, and since a Pauli string not present in $H$ may be included as long as they cancel out in the overall sum \cite{choi2022ghost,choi2023fluid}. Finding the optimal partitioning is a hard problem and we introduce heuristic methods to find a reasonable partitioning.

\subsection{Illustrative example}

Now that we have introduced the general method, let us know consider an example to illustrate the idea and advantage of the approach. Consider the following (artificial) four-qubit Hamiltonian
\begin{equation}
\begin{split}
H = I +
\frac{1}{2} X_0 X_1 X_2 +
X_0 X_2 +
\frac{1}{2} Y_0 Y_1 X_2 +
X_1 + 
2 X_1 X_2 + \\
 X_1 X_2 X_3 
- X_1 X_2 Z_3 
- X_1 Z_2  
-\frac{1}{2} X_1 Z_2 X_3 + \\
\frac{1}{2} X_1 Z_2 Z_3 +
\frac{1}{2} X_1 X_3  
-\frac{1}{2} X_1 Z_3 + 
4 X_2 + 
 X_2 X_3  \\
- X_2 Z_3 
- Z_2  
-\frac{1}{2} Z_2 X_3 + 
\frac{1}{2} Z_2 Z_3 +
\frac{1}{2} X_3 
-\frac{1}{2} Z_3.
\end{split}
\end{equation}
Implementing the SortedInsertion \cite{crawford2021si} algorithm with full commutation yields five sets of mutually commuting Pauli terms:
\begin{equation}
\begin{split}
\{X_2 ,
X_1 X_2 ,
X_2 X_3 ,
X_1 X_2 X_3 ,
X_1 ,
X_0 X_2 ,
X_3 ,
X_1 X_3 , \\
X_0 X_1 X_2\};
\{Z_2 ,
X_1 Z_2 ,
Z_2 X_3 ,
X_1 Z_2 X_3\}; \\
\{Z_2 Z_3 ,
X_1 Z_2 Z_3\}; \\
\{X_2 Z_3 , 
X_1 X_2 Z_3 ,
Z_3 ,
X_1 Z_3\}; \\
\{Y_0 Y_1 X_2\}.
\end{split}
\end{equation}
Hence one can evaluate $\expval{H}$ using \textit{five} measurement bases. Next we consider \knoclid{}. With 2-\noclid, there exists the following decomposition into just \textit{two} measurement bases $M_1 = W_{11} + W_{12}$ and $M_2 = W_{21}$, where
\begin{align}
\begin{split}
W_{11} &= 
\begin{pmatrix}
1 & 0 & 1 & 0 \\
0 & 1 & 0 & 1 \\
1 & 0 & 1 & 0 \\
0 & 1 & 0 & 1 \\
\end{pmatrix}
\otimes
\begin{pmatrix}
0 & 1 \\
1 & 1 \\
\end{pmatrix}
\otimes
\begin{pmatrix}
0 & 1 \\
1 & 2 \\
\end{pmatrix}
\end{split}
\nonumber \\
W_{12} &= 
\begin{pmatrix}
1 & 0 & 1 & 0 \\
0 & 1 & 0 & 1 \\
1 & 0 & 1 & 0 \\
0 & 1 & 0 & 1 \\
\end{pmatrix}
\otimes
\begin{pmatrix}
0 & 1 \\
1 & 1 \\
\end{pmatrix}
\otimes
\begin{pmatrix}
1 & 0 \\
0 & 1 \\
\end{pmatrix}
\\
W_{21} &= 
\begin{pmatrix}
2 & 1 & 0 & 0 \\
1 & 2 & 1 & 0 \\
0 & 1 & 2 & 1 \\
0 & 0 & 1 & 2 \\
\end{pmatrix}
\otimes
\begin{pmatrix}
0 & 1 \\
1 & 0 \\
\end{pmatrix}
\otimes
\begin{pmatrix}
1 & 0 \\
0 & 1 \\
\end{pmatrix}.
\nonumber
\end{align}
Although, as mentioned, the number of measurement bases does not always correlate with the number of shots needed, this example illustrates how relaxing Clifford constraints in \knoclid{} yields fewer measurement bases, which could potentially decrease the number of shots. Indeed, in our numerical results below we directly study the effects of this new partitioning scheme on shot counts and show an advantage in \knoclid{} in several applications of interest.

\subsection{Hamiltonian classes}

Here we summarize the four Hamiltonian classes considered in this work. They include problems with either bosonic or fermionic commutation relations. Though the methods we discuss here are general to any quantum operator, we introduce these Hamiltonian classes here because we discuss many of our methods explicitly in terms of these common Hamiltonians.

The spinless Fermi-Hubbard \cite{abrams1997simulation,cade2020strategies,leblanc2015solutions} lattice model is defined as
\begin{equation} \label{eqn:spinlessfh}
H = - \sum_{\langle i, j \rangle} t_{ij} (a^\dagger_i a_j + a^\dagger_j a_i)
    + \sum_{\langle i, j \rangle} \frac{U_{ij}}{2} a^\dagger_i a_i a^\dagger_j a_j
\end{equation}
where $i$ and $j$ denote lattice sites, $t_{ij}$ are hopping terms, and $U_{ij}$ are repulsion terms. We consider only models with nearest-neighbor interactions.

The molecular electronic structure Hamiltonian is defined as
\begin{equation}\label{eq:hamelec}
H = \sum_{pq} t_{pq} a_p^\dag a_q + \half \sum_{pqrs} t_{pqrs} a_p^\dag a_q^\dag a_r a_s.
\end{equation}
where orbitals are labelled $p,q,r,s$. In this work, for both Fermi-Hubbard and electronic structure, we implement only the Jordan-Wigner transformation \cite{whitfield2011simulation,babbush2015chemical}.

The Bose-Hubbard model \cite{elstner1999dynamics,somma2003quantum,bahrami2024bempa} describes bosonic particles on a lattice,
\begin{equation}\label{eq:bh}
H_{BH} = - \sum_{\langle i,j \rangle} t_{ij} \left(  b_{i}^\dag b_j + h.c.  \right) +  \sum_i  \frac{U_i}{2}  n_i (n_i - 1)
\end{equation}
where $i$ and $j$ denote lattice sites, $t_{ij}$ are hopping terms, and $U_i$ is the on-site interaction. 

Many chemical processes and properties depend on the behavior of the nuclei, as described by a molecular vibrational Hamiltonian  \cite{mcardle2019digital,ollitrault2020hardware,jnane2021noncondon,sawaya2023notrap}. One uses standard bosonic operators to produce a Hamiltonian
\begin{equation}\label{eq:ham_aharm}
\begin{split}
H_{v} = \half \sum_i^M \omega_{i}(q_i^2 + p_i^2) \\
+ \sum_{\{ijk\}} t_{ijk} q_iq_jq_k   \\
+ \sum_{\{ijkl\}} t_{ijkl} q_iq_jq_kq_l + \cdots,
\end{split}
\end{equation}
where indices label the vibrational mode and the expansion is to arbitrarily high order. Though we use the harmonic basis, other bases may be used. While $H_v$ is classically solvable for some molecules, in general the anharmonic coupling terms lead to this being a hard problem, as discussed elsewhere \cite{sawaya2021pra}.

A notable property of bosonic and vibrational problems is their canonical bosonic commutation relations, because (unlike in the fermionic case) this leads to qubit-encoded Hamiltonians that have the same tensor-product structure as the bosonic Hamiltonian. As we discuss below, this allows one to use tensor products of $\{q_i\}$, $\{p_i\}$, and $\{n_i\}$ as the measurement basis for \textit{any} bosonic or phononic problem instance. For encoding \cite{somma2005quantum, mcardle2019vibr, ollitrault2020hardware, sawaya2022dqir} bosonic and phononic modes we used the Gray code \cite{sawaya2020resource,sawaya2020connectivity}.

We note that expectation value estimation is a required subroutine for some near-term quantum algorithms for classical problems as well \cite{jiang2015quantum,lubasch2020variational,larose2020robust,schuld2021effect,xu2021variational,amankwah2022quantum,jumade2023data,bravo2023variational}. However, for this work we focus entirely on quantum problems.

\subsection{Partitioning via greedy algorithm}\label{sec:greedy}

Here we describe the first of our \knoclid{} partitioning methods, a simple greedy algorithm that operates on the encoded (qubit space) Hamiltonian. Pseudocode for the greedy algorithm is given here:

\vspace{1em}
\begin{algorithmic}
\State $\vartriangleright$ FREEQUBITS() returns the number of mismatched qubits, where the local Pauli is different on different qubits.
\State $\vartriangleright$ DISJOINT() returns true only if the inputs operate on disjoint qubits.
\For{$P_i$ in $\{P_i$ ordered by $|c_i|$ descending\} }
    \For{ $M_q$ in $\{M_q\}$ }
        \For{ $W_{qr}$ in $\{W_{qr}\}$ }
            \If{ FREEQUBITS($W_{qr}$,$P_i$) $\leq k$ }
                \State $W_{qr} = W_{qr} + P_i$
                \State \textbf{break} 2 loops
            \EndIf
            \If{ $P_i$ rejected from all $W_{qr}$ }
                \If{ DISJOINT($W_{qr}$,$P_i$) for all $W_{qr}$ }
                    \State Initialize new $ W_{q,r+1}=P_i$
                    \State \textbf{break} 2 loops
                \EndIf
            \EndIf
        \EndFor
    \EndFor 
    \If{$P_i$ rejected from all $\{M_q\}$}
        \State Initialize new $M_{q+1}$
        \State Add $W_{q+1,1}=P_i$ to $M_{q+1}$
    \EndIf
\EndFor
\end{algorithmic}
\vspace{1em}

The basic idea is to loop through terms of the qubit-encoded operator, determining how many mismatched qubits are present in a given $W_{qr}$. These (at most $k$) ``free qubits" are the ones across which a diagonalization unitary will be placed. 
If an attempt to place a term $P_i$ into $W_{qr}$ leads to more than $k$ mismatched qubits, then the algorithm attempts to form a new $W_{q,r+1}$ in partition $M_q$. This attempt succeeds only if $W_{qr}$ and $P_i$ operate on disjoint qubits; otherwise, an attempt is made to place $P_i$ into $M_{q+1}$. Only after placement attempts fail for all $M_q$ is a new $M_{q'}$ created.

The $P_i$ are looped through in descending order of importance, by magnitude of $c_i$, as in the SortedInsertion approach \cite{crawford2021si}. Note, requiring that $W_{qr}$ and $W_{qr'}$ (with $r \neq r'$) operate on disjoint qubits is a significant restriction that was imposed only for ease of implementation; removing this restriction would lead to fewer partitions.

\subsection{Blocking with residuals}\label{sec:blocking}

Here we introduce a second partitioning method that also operates directly on the qubit-encoded Hamiltonian. The main idea is to partition the qubits into sets of size at most $k$, and then pack as many operators as possible into those qubit sets. After several iterations, the remaining unassigned terms (which we call the residual) in the qubit Hamiltonian, which will tend to be longer Pauli strings, are treated with a separate method. 

For the blocking method we define an arbitrary number of bases $M_q^{\text{B}}$. These are constructed in ascending order ($M_0^{\text{B}}$ before $M_1^{\text{B}}$, etc.). $M_i$ is constructed by defining a disjoint collection of sets, where each set $\m K_q\ep{1}, \m K_q\ep{2}, \dots$ contains at most $k$ qubit indices 
\begin{equation}
M_q^{\text{B}} = C_{q}^{\m K_q\ep{1}} \otimes C_{q}^{\m K_q\ep{2}} \otimes C_{q}^{\m K_q\ep{3}} \otimes ...
\end{equation}
such that any $C_{q}^{\m K_q\ep{s}}$ contains all remaining terms that operate exclusively on qubits in $\m K_q\ep{s}$ (``remaining'' because some operators may have already been placed in a previous $M_j$ with $j<q$).

Though the only constraint is that $|\m K_i|<k$ for all $i$, for this work we use contiguous sets of qubits for this blocking algorithm (this is not the case for our greedy algorithm). 
For $M_1^{\text{SS}}$, we use $\mathcal K_1=\{1,\cdots,k\}, \m K_1=\{k+1,\cdots,2k\}, \cdots$; for $M_2^{\text{SS}}$, we use $\mathcal K_2=\{2,\cdots,k+1\}, \m K_1=\{k+2,\cdots,2d+1\}, \cdots$; and so on. This leads to $k$ bases $M_i$.

This procedure may leave a significant number of remaining terms, especially in the case of electronic structure. For example, any term that operates on qubits more than $k$ indices apart will not be placed into any of the $M_q$ partitions. Hence the remaining terms are partitioned using a different method; our implementation uses SortedInsertion with full commutation \cite{crawford2021si} on these remaining terms.

\subsection{Index reordering}\label{sec:reorder}

In electronic structure, the Jordan-Wigner transform yields long Pauli strings (\textit{i.e.} terms with high Pauli weight). Because it will typically be easier to fill a partition $M_q$ with many low-weight terms than with many high-weight terms, it will often be beneficial to re-order the electronic Hamiltonian \eqref{eq:hamelec}. Intuitively, one would like to pack the more significant terms (those with larger $|t_{pq}|$ and $|t_{pqrs}|$) into as few partitions $M_q$ as possible; hence one would like such terms with larger $|h|$ to have smaller Pauli weight. To this end we use the following cost function to optimize index ordering:
\begin{equation}
\begin{split}
C_\text{ordering} = \sum_{pq} \| t_{pq} \| \|p-q\| \\
+ \sum_{pqrs} \| t_{pqrs} \| \left( \text{max}(p,q,r,s) - \text{min}(p,q,r,s) \right).
\end{split}
\end{equation}
Optimizing with respect to this cost function ensures that, after the Jordan-Wigner transform, the more significant terms span fewer qubits. We implemented a simple reindexing algorithm as follows. First we swap a random pair of indices. The swap is kept only if $C_{ordering}$ decreases; otherwise the indices are placed in their previous position. In our implementation, the algorithm is halted after some threshold of failed swaps in a row.

\subsection{Partitioning via pre-encoded operators}\label{sec:preencoded}

Methods (such as our simple greedy algorithm above) that operate directly on the qubit space are unlikely to take advantage of much of the structure of the ``pre-encoded'' (\textit{i.e.} expressed as bosonic or fermionic operators) Hamiltonian. In a sense this structure is lost or hidden after the problem is encoded to qubits. It may instead be useful to explicitly consider the pre-encoded operator, in order to directly take its particular tensor product structure into account. 
Let us assume that the pre-encoded Hamiltonian is known and is expressible as 
\begin{equation}\label{eq:preencoded}
H = \sum_i B_{i0}^{[k_{i0}]} \otimes B_{i1}^{[k_{i1}]} \otimes B_{i2}^{[k_{i2}]} \otimes \cdots
\end{equation}
where superscripts denote the logarithm of the size of each tensor. Then a greedy search algorithm may not be needed for \knoclid{}, as it may be more straightforward to use the components of \eqref{eq:preencoded} to intuitively construct the $W_{qr}$ of equation \eqref{eq:ttrain}. The benefits of this strategy are most obvious with Hamiltonians of bosonic (or phononic) degrees of freedom. The following three sections consider the structure of the pre-encoded operators when implementing a partitioning plan.

\subsection{Lattice edge coloring}\label{sec:coloring}

In lattice models, we propose using an edge coloring approach to arrive at a \knoclid{} ~factorization. Define $\mathcal E$ to be the set of all edges in the lattice.

In the Bose-Hubbard model, in each edge coloring one can simultaneously measure all $b_i^\dag b_j + b_i b_j^\dag$ where edge $(i,j)$ belongs to the same color. An edge coloring leads to sets of edges denoted $\m E_1, \m E_2, \cdots, \m E_{n_c}$ where $n_c$ is the number of colors and $\bigcup_{k} \mathcal E_k = \mathcal E$. Then one measures in $n_c$ bases
\begin{equation}\label{eq:hopbasis}
M_k^{\text{BH,hop}} = \sum_{\{(i,j\}) \in \m E_q} \left( b_i^\dag b_j + b_i b_j^\dag \right)
\end{equation}
where $\{i,j\}$ is an edge in the set $\mathcal E_k$. One also measures in the diagonal $Z$ basis to recover all $n_i(n_i-1)$ terms in equation \eqref{eq:bh}, leading a total of just $n_c + 1$ bases in which to measure. $M_k^{\text{BH,hop}}$ requires diagonalization across two bosonic modes, \textit{i.e.} the diagonalization parameter $k=2 \log_2 d$. (In the next subsection we give a method that allows $k$ to be halved.) We summarize well-known results \cite{kattemolle2024edge} for colorings of regular lattices in Table \ref{tbl:edgecoloring}.

Now we consider the one-dimensional spinless Fermi-Hubbard model with only nearest-neighbor interactions. Fermionic case 1D. Regardless of system size, only two 2-\noclid ~bases are required:
\begin{equation}\label{eq:fh_color}
\begin{split}
M_0^{\text{FH,1D}} &= 
  (t X_0X_1 + t Y_0Y_1 - \frac{U}{8} Z_0Z_1 + \frac{U}{8} Z_0) \\ 
&+ (t X_2X_3 + t Y_2Y_3 - \frac{U}{8} Z_2Z_3 + \frac{U}{8} Z_2) + \cdots \\
M_1^{\text{FH,1D}} &= 
  (t X_1X_2 + t Y_1Y_2 - \frac{U}{8} Z_1Z_2 + \frac{U}{8} Z_1) \\ 
&+ (t X_3X_4 + t Y_3Y_4 - \frac{U}{8} Z_3Z_4 + \frac{U}{8} Z_3) + \cdots
\end{split}
\end{equation}
Hence one would diagonalize the 2-qubit operator $(t X_iX_j + t Y_iY_j - \frac{U}{8} Z_iZ_j + \frac{U}{8} Z_i)$ simultaneously across $(i,j) \in ((0,1), (2,3), \cdots)$, and then simultaneously across $(i,j) \in ((1,2), (3,4), \cdots)$. Contrast this with commuting grouping methods, where one would need 3 bases: $XX+YY$ for two colors, and then $Z$ for all sites. This smaller number of bases (two instead of three) is not guaranteed to produce lower shot counts, but it does provide one more option for basis measurement. A more complicated coloring scheme is also possible for higher-dimensional Fermi-Hubbard models. We summarize the number of colors (which is equivalent to the number of bases) for different lattice types in Table \ref{tbl:lattices}, but we leave further details for future work.

\begin{table}[h!]
\label{tbl:edgecoloring}
\centering
\begin{tabular}{|c|c|c|c|c|c|c|}
\hline
\textbf{ } & \textbf{1D} & \textbf{Square} & \textbf{Hex} & \textbf{Triang.} & \textbf{Cubic} & \textbf{Tetrahed.} \\
\hline
\textbf{Colors} & 2 & 4 & 3 & 6 & 6 & 4  \\
\hline
\hline
\end{tabular}
\caption{Number of colors in optimal edge coloring of different lattice types.}
\label{tbl:lattices}
\end{table}

\subsection{Products of one-body Hermitians}\label{ssec:sumshermitian}

In the previous section we introduced a lattice coloring scheme, because hopping terms $b_i^\dag b_j + b_i b_j^\dag$ on adjacent edges cannot be simultaneously measured. Here we introduce a method to avoid coloring altogether, which results in a much lower (and in fact constant) number of measurement bases. The key insight is that the following equality holds:
\begin{equation}\label{eq:qqpp}
b_{i}^\dag b_j + h.c. = q_i q_j + p_i p_j.
\end{equation}
The RHS is useful because (unlike the LHS) it is \textit{a sum of products of one-body Hermitians}. (Such constructs are useful in other contexts in quantum algorithms for electronic structure \cite{motta2023bridging}.) Because the one-body non-Hermitian $\expval{a_i^\dag}$ and $\expval{a_i}$ cannot be directly measured on a quantum computer, it would appear that the entire two-body hopping term $\expval{b_{i}^\dag b_j + h.c.}$ must be measured, which is why diagonalization must be across $k=2 \log_2 d$ qubits.

But if we use the RHS of equation \eqref{eq:qqpp}, we can simply measure $q$ in \textit{all} bases (with diagonalization across only $k=\log_2 d$ qubits), followed by $p$ in all bases, leading to measurement basis operators
\begin{equation}
\begin{split}
M_q^{\text{BH,quad}} &= \sum_{\{(i,j\}) \in \m E} q_i q_j \\
M_p^{\text{BH,quad}} &= \sum_{\{(i,j\}) \in \m E} p_i p_j \\
M_n^{\text{BH,quad}} &= \sum n_i. 
\end{split}
\end{equation}
This is massively beneficial because we can now both (a) reduce $k$ by half and (b) reduce the number of measurement bases to a constant of just three. This means we can measure in three bases \textit{regardless of lattice type}, even an all-to-all lattice. Note that $M_n^{\text{BH,quad}}$ can be measured in the Z-basis without additional diagonalizatoin circuit.

In the structure of vibrational Hamiltonians in the form of equation \eqref{eq:ham_aharm}, the same strategy may be used. Analogously to the Bose-Hubbard case, one can measure a vibrational Hamiltonian using only \textit{two} bases: ${q_1, q_2, \cdots}$ and ${p_1, p_2, \cdots}$. 
\begin{equation}
\begin{split}
M_q^{\text{vibr}} &= \half \sum_i^M \omega_{i}q_i^2 + \sum_{\{ijk\}} c_{ijk} q_iq_jq_k   \\
&+ \sum_{\{ijkl\}} c_{ijkl} q_iq_jq_kq_l + \cdots \\
M_p^{\text{vibr}} &= \half \sum_i^M \omega_{i} p_i^2 \\
\end{split}
\end{equation}
Each $A_{qrs}^{\m K_{qrs}}$ in equation \eqref{eq:ttrain} is simply a $q$ or $p$, and all $q$-containing tensor products in equation \eqref{eq:ham_aharm} can be measured simultaneously, regardless of order or of number of terms. Hence any vibrational Hamiltonian of form \eqref{eq:ham_aharm} may be measured in just \textit{two} bases.

\subsection{Sub-operators}\label{sec:subops}

In previous sections we have implicitly required diagonalization circuits over $k$ qubits, where $k$ is often determined by the problem of interest (mainly via the truncation of the bosonic or phononic modes). This is allowable because $k$ is small for many if not most applications: low-temperature Bose-Hubbard simulations usually can use truncations less than $d=8$ ($k$=3), and many molecules of interest do not require vibrational cutoffs of more than $d=16$ ($k$=4). Hence we do not often expect circuit depths to be an obstacle. However, there will be cases where a large $k$ requires circuit depths that are too high.

In such cases, one needs to split the original $k'$-local operators into tensor decompositions with tensor size $k < k'$. We decompose the tensors in equation \eqref{eq:ttrain} as
\begin{equation}
\begin{split}
A_{qrs} \rightarrow \sum_{l} \bigotimes_m \hat a_{qrslm} \\
\end{split}
\end{equation}
such that all $\hat a$ operate on $\leq k'$ qubits. This construction is more relevant to bosonic and phononic Hamiltonians, because in the fermionic case one tunes $k$ using unrelated methods as discussed below. %

This sub-operator decomposition allows one to use the methods of the previous two sections, while still imposing arbitrarily small $k$. Now that one is measuring each term in the new decomposition separately, an edge coloring is again required even for the one-body Hermitian string strategy, which was not that case if $k'=k$. We leave a more thorough treatment of this sub-operator strategy to future work.

\section{Numerical results}

Here we present numerical results for shot counts, for four Hamiltonian classes. Note that in the interest of rigor, we calculate exact shot counts $N$, circumventing useful but order-of-magnitude approximations like the $\hat R$ metric \cite{crawford2021si}. Hence our simulations do not exceed 16 qubits. We run the algorithms on random quantum states and report results in terms of $\varepsilon^2 N$ (see~\eqref{eq:generalshotcounts}), which we refer to as the measurement variance. Where index reordering was performed, the algorithm was halted after 5000 failed swaps in a row. Hamiltonians were taken from the HamLib dataset \cite{hamlib,butko2024hamperf} or produced using software packages OpenFermion \cite{mcclean2020openfermion} and mat2qubit \cite{sawaya2022mat2qubit}.

QWC-SI and FC-SI denote qubit-wise commutation and full commutation, both using SortedInsertion as the partitioning algorithm. Lower bounds are calculated by diagonalizing the full Hamiltonian; in other words, implementing formula \eqref{eq:generalshotcounts} with a single fragment. Greedy, Blocked, and Coloring denote the methods of Sections \ref{sec:greedy}, \ref{sec:blocking}, and \ref{sec:coloring}, respectively. QPN and QP denote the method of Section \ref{ssec:sumshermitian}, using measurement bases $\{q_i,p_i,n_i\}$ and $\{q_i,p_i\}$, respectively.

\begin{figure}\label{fig:fh_res}
    \centering
    \includegraphics[width=1.0\linewidth]{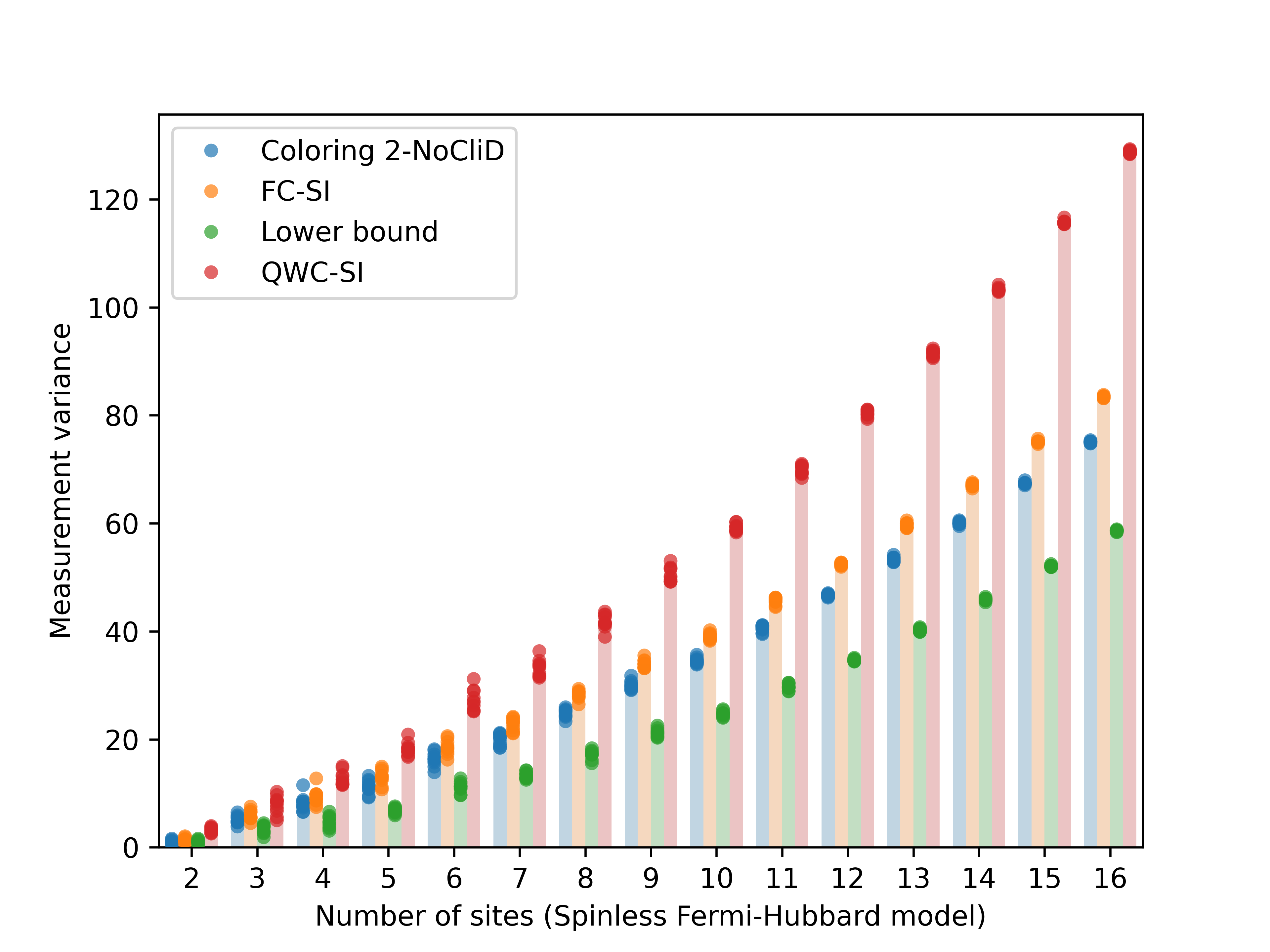}
    \caption{Measurement variance $\varepsilon^2 N$ for the spinless Fermi-Hubbard model.}
    \label{fig:fhres}
\end{figure}

\begin{figure}\label{fig:bh_res}
    \centering
    \includegraphics[width=1.0\linewidth]{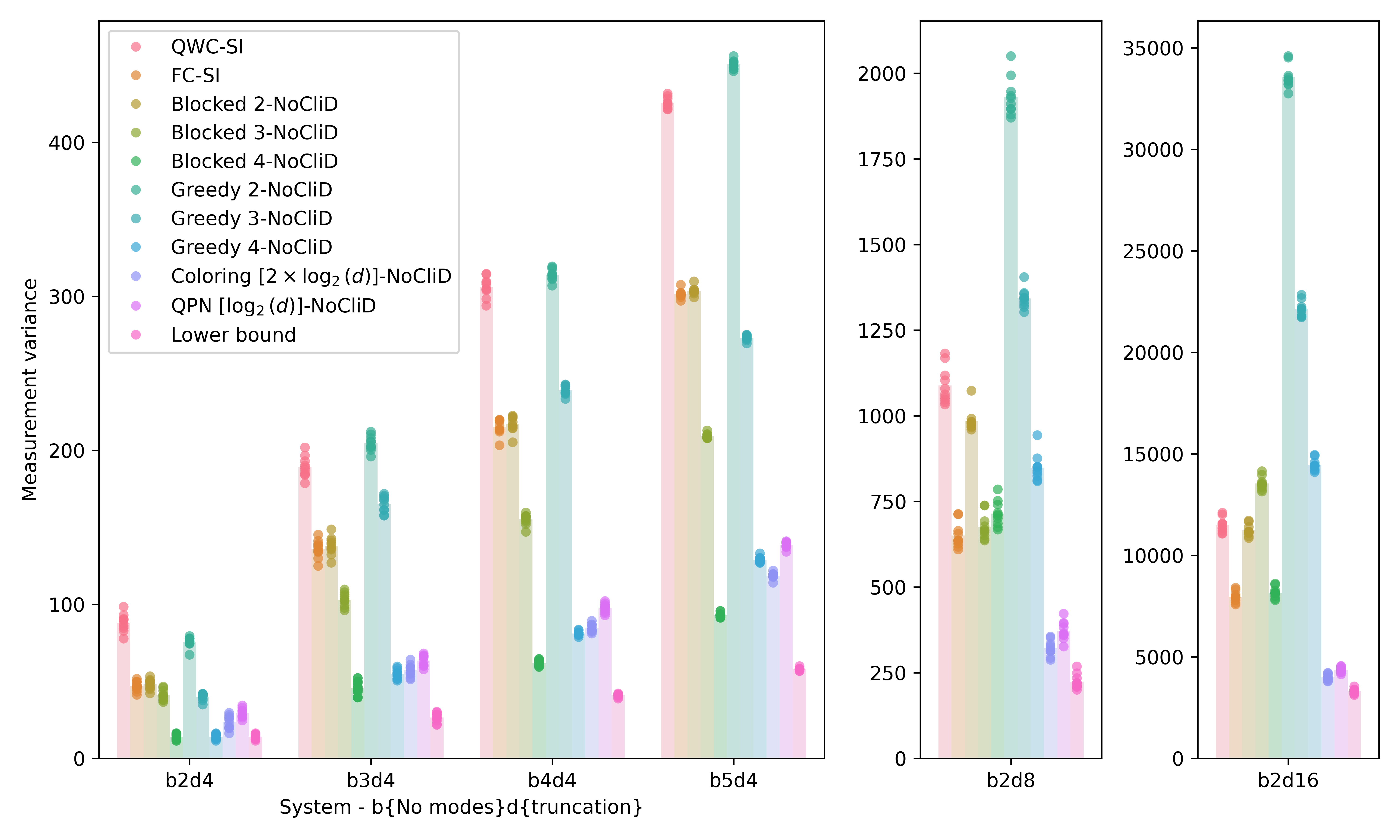}
    \caption{Measurement variance $\varepsilon^2 N$ for the Bose-Hubbard model. Labels denote the number of bosonic modes and the bosonic truncation. For example, b3d4 signifies 3 bosonic modes with a truncation of $d=4$ in each mode.}
    \label{fig:bhres}
\end{figure}

\begin{figure*}\label{fig:vibr_res}
    \centering
    \includegraphics[width=1.0\linewidth]{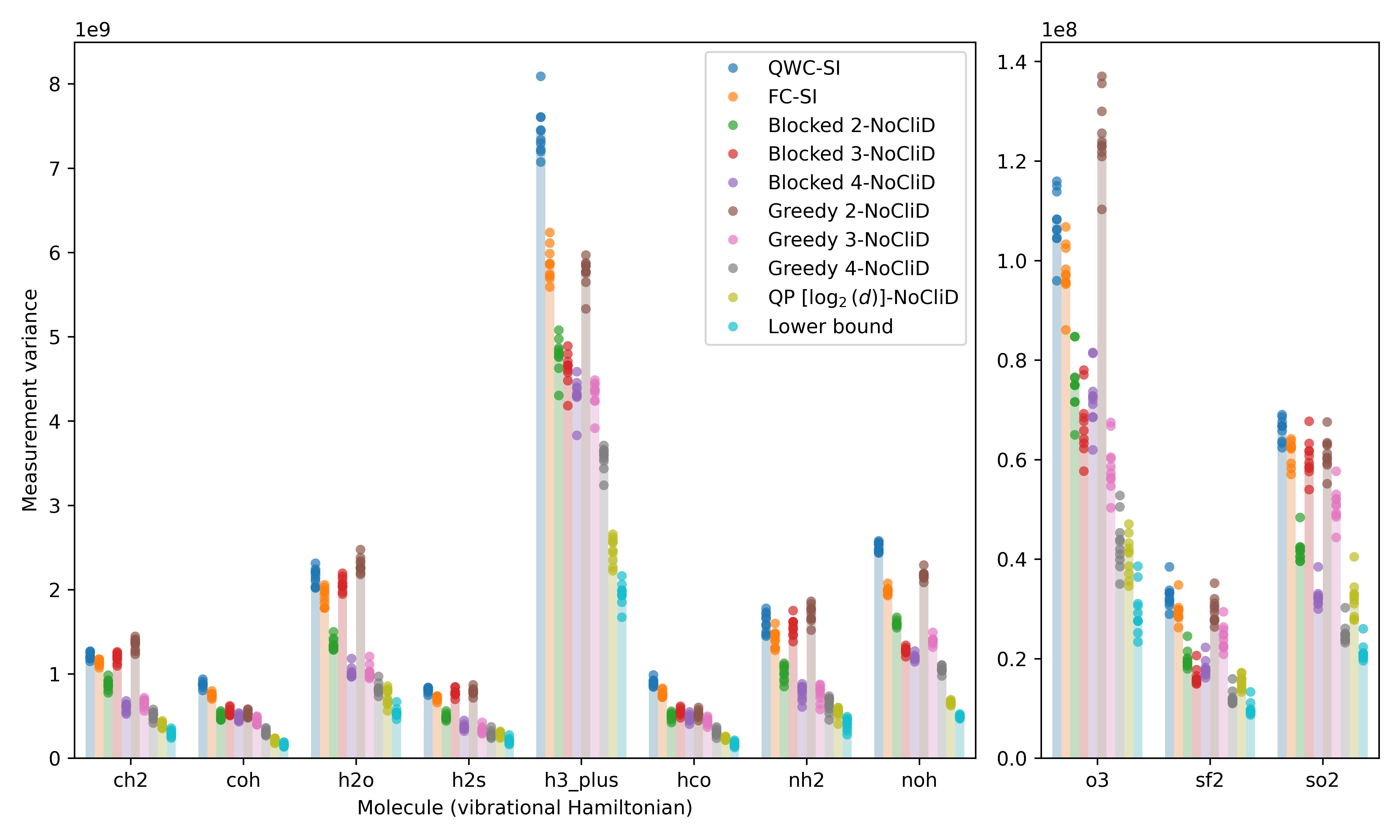}
    \caption{Measurement variance $\varepsilon^2 N$ for molecular vibrational structure in units cm$^{-2}$, where all modes in the model are truncated as $d=4$.}
    \label{fig:vibrres}
\end{figure*}

For the spinless Fermi-Hubbard model (Figure \ref{fig:fhres}), the NoCliD-coloring method out-performs SortedInsertion, though they are comparable in magnitude. However, that these results are for $k=2$ as in equation \eqref{eq:fh_color}, shows that the improvement in shot counts is achieved at negligible cost in circuit depth.

For electronic structure, because of the higher Pauli weights, we do not expect these rudimentary partition algorithms to easily out-perform FC-SI. However, we were interested in demonstrating \textit{some} threshold value of $k^*$ beyond which \noclid{} leads to a shot count lower than FC-SI. We found $k^*$=6 for LiH, $k^*$=7 for HF, NH, and NaLi, $k^*$=8 for BH, C2, F2, H6, O2, and OH, $k^*$=9 for B2, BeH, CH, and Na2, and $k^*$=10 for N2, where Hamiltonians were taken from HamLib. In these cases, unlike in the bosonic cases, NoClid-blocking performed slightly worse than NoClid-greedy. 
We stress that the ``disjoint'' constraint of Section \ref{sec:greedy}'s greedy algorithm is extremely restrictive, likely leading to far more partitions than necessary. This threshold for $k$ is likely to be significantly reduced in electronic structure Hamiltonians by relaxing this constraint, and we leave this as a direction for future work.

Across all considered instances of Bose-Hubbard (Figure \ref{fig:bhres}), \knoclid{} methods consistently outperform traditional methods. When comparing coloring and QPN methods, QPN is slightly superior in most cases. Given that QPN achieves this performance with $k$ values half as large, it is recommended as the preferred method. Additionally, methods like greedy and blocking, which do not account for the original Hamiltonian structure, are inferior to both coloring and QPN approaches. This is unsurprising, as the latter methods explicitly consider the bosonic structure of the problem.

Finally, for vibrational structure (Figure \ref{fig:vibrres}), all NoCliD methods outperform the traditional SortedInsertion approach. Similar to the Bose-Hubbard case, the blocking and greedy methods are easily outperformed by the QP method, as they fail to account for the bosonic structure. The QP method also performs very close to the theoretical lower bound, which is perhaps unsurprising as it requires only two bases regardless of the Hamiltonian instance.  These results are nearly optimal as they are very close to the lower bound of diagonalizing the full Hamiltonian, vastly out-performing all methods. This is an especially notable results considering that $k$ is very small ($k=2$) for these QP results.

In all four Hamiltonian classes, at least one \knoclid{} partitioning strategy led to smaller measurement variance with respect to SortedInsertion with qubit-wise commutation (QWC) and full commutation (FC). The largest improvements were observed for bosonic and phononic degrees of freedom even for very small $k$, partly because these Hamiltonians tend to have shorter Pauli products in qubit space and partly because it is easier to tailor \knoclid{} to these Hamiltonians than to electronic structure. For large enough $k$, \knoclid{} leads to fewer shots even for electronic structure, though future work is needed to reduce this threshold value of $k$.

\section{Conclusions}

We introduced a class of methods for reducing shot counts in expectation value estimation. Our approach, referred to as $k$-local non-Clifford diagonalization (\knoclid{}), allows for partitioning of arbitrary terms, as long as the size of each tensor in the tensor product operates on at most $k$ qubits. As the basis-change circuit depth increases with $k$, the method allows for a natural trade-off between shot counts and circuit depth. We calculated exact shot counts required for four Hamiltonians classes: Bose-Hubbard, Fermi-Hubbard, vibrational structure, and electronic structure. In all cases, there is some value of $k$ for which \knoclid{} out-performs other partitioning methods based on mutually commuting partitions.

For now, the method's utility is clearest for vibrational, bosonic, and Fermi-Hubbard Hamiltonians, where the method often out-performs fully-commuting methods by $k$ as small as 2 (for which the basis change circuit is extremely shallow). These positive results are largely due to the fact that we developed specialized partitioning algorithms for these Hamiltonians. For electronic structure, we left the development of more sophisticated partitioning methods to future work, which we expect to drastically reduce the threshold value of $k$. Other future research directions include extensions of the method based on `ghost' \cite{choi2022ghost} and `fluid' \cite{choi2023fluid} partitioning methods introduced previously, as well as a thorough study of the basis-change quantum circuits.

\section*{Acknowledgements}

This research used resources of the National Energy Research Scientific Computing Center (NERSC), a Department of Energy Office of Science User Facility using NERSC awards ASCR-ERCAP0029116, DDR-ERCAP0030342, and DDR-ERCAP0033798. This work was supported by Wellcome Leap as part of the Quantum for Bio Program.

\appendix

\section{Valid and invalid tensor products for \knoclid{}}

In order to build some intuition for allowable fragments in \knoclid{}, here we show some examples of valid Pauli string matchings for a given $k$. ``Matched'' local operators are denoted with blue. %

For k=1, an example of a valid string matching set is \{ $a$\texttt{XX\blue{Z}}, $b$\texttt{XX\blue{Y}} \}. Another example is \{ $a$\texttt{I\blue{Z}I}, $b$\texttt{I\blue{X}I} \}. That diagonalizing qubit is denoted in blue. Note that unlike in commuting set methods, in the general case real coefficients $a$, $b$, ... need to be included in the defition; though this is not stricly required for the basic versions of \knoclid{}, it will be required for the more complicated factoring methods discussed below. 

For k=2, examples of valid string matchings are \{ $a$\texttt{IXX\blue{ZI}}, $b$\texttt{IXX\blue{XY}} \} and \{ $a$\texttt{I\blue{Z}I\blue{Z}I}, $b$\texttt{I\blue{X}I\blue{I}I} \}. Note that the local operator on black qubits must match exactly, while there is complete freedom for the blue qubits. 

The point is, the overall operator must be able to be factored into a tensor product of operators as in equation \eqref{eq:ttrain}, where each operator in the tensor product operates on at most $k$ qubits.

\textbf{Non-examples.} Here are some examples of \textit{invalid} string-matching. Some non-examples of partitionings for k=1 are \{ $a$\texttt{\gray{I}XX\gray{Z}}, $b$\texttt{\gray{X}XX\gray{Y}} \} and even \{ $a$\texttt{\gray{I}XX\gray{Z}}, $b$\texttt{\gray{X}XX\gray{I}} \}, where non-identical local operators are in gray (these are valid $k=2$ matchings but invalid $k=1$ matchings). It is important to consider the latter set; it shows that even if a set is $k$-qubitwise-commuting, that does not make it a valid \knoclid{} matching. This highlights the fact that, unlike in the case of commuting grouping, \knoclid{} indeed requires that the operator be representable as equation \eqref{eq:ttrain} such that all $k_i < k$. %

For $k$=1, another example of an \textit{invalid} string matching is \texttt{XI\textbf{X}} and \texttt{IX\textbf{Z}}. Even though the first two qubit qubit-wise commute, the set cannot be expressed as a product of tensors each operating on at most 1 qubit.

\section{Proof of 1-qubit basis change result}\label{apx:thm1q}

Before proceeding to the proof of Result 1, we begin this section with two examples in order to build intuition.

Consider the following pedagogical Hamiltonian, 
\begin{equation}\label{eq:H1}
H_1 = \roothalf X + \roothalf Z.
\end{equation}
We refer to the basis in which equation \eqref{eq:H1} is defined as the global Pauli basis (GPB).

We define $P_{O} \equiv \mathcal N O$ for any one-qubit Hermitian operator $O$, where $\mathcal N$ is a normalization constant such that $P_{O}^2 = I$. In other words, $P_{O}$ is a rotated Pauli matrix. For Hamiltonian $H_1$, we use
\begin{equation}
\begin{split}
P_{X+Z} = (X+Z)/\sqrt{2}
\end{split}
\end{equation}
The point is to measure in the $P_{X+Z}$ basis directly instead of measuring in the $X$ or $Z$ Pauli bases. For one qubit, this entails simply rotating on the Bloch sphere to an orientation halfway between the $X$ and $Z$ axes. We note that, by definition, the following properites are true for all $P_O$:

\begin{equation}
\begin{split}
\text{Var}[P] &= \expval{P^2} - \expval{P}^2 \\
\text{Var}[P] &= 1 - \expval{P}^2
\end{split}
\end{equation}

Now we consider the total number of measurements required to calculate $\expval{H_1}$ for two distinct states:  $\ket{\phi}=R_y(\pi/4)\ket{0}$ $\approx [0.924,0.383]^T$ and $\ket{0}$. For each state we consider measuring in (a) the global Pauli basis (GPB) (which is restricted for measuring only $\ev{X}$ and $\ev{Z}$) and (b) the $\braket{P_{X+Z}}$ basis which we refer to as the rotated basis (RB).

Note that
\begin{equation}
R_y(\theta) = 
\begin{bmatrix}
\cos(\theta/2) & -\sin(\theta/2) \\
\sin(\theta/2) &  \cos(\theta/2)
\end{bmatrix}
\end{equation}

Measuring $\ev{A}{\phi}$ in the global Pauli basis yields:
\begin{equation}
\begin{split}
\varepsilon^2 N_{tot}^{GPB} &= \left( \sqrt{\Var{\roothalf X}} + \sqrt{\Var{\roothalf Z} } \right)^2 \\
&= \left(  \sqrt{\half(1-\ev{X}{\phi}^2)} + \sqrt{\half(1-\ev{Z}{\phi}^2)} \right)^2 \\
&= \left(  \sqrt{\half(1-\half)} + \sqrt{\half(1-\half)} \right)^2  \\
&= 1
\end{split}
\end{equation}

Measuring $\ev{A}{\phi}$ in the RB basis $P_{X+Z}$ yields:
\begin{equation}
\begin{split}
\varepsilon^2 N_{tot}^{RB} &=  \Var{\ev{P_{X+Z}}{\phi}}  \\
&= ( 1 - \ev{P_{X+Z}}{\phi}^2 ) \\
&= ( 1 - 1 ) \\
&= 0 < \veps^2 N_{tot}^{GPB}
\end{split}
\end{equation}

So for this example state $\ket{\phi}$ it is indeed beneficial to move to the rotated basis. This is expected, as $\ket{\phi}$ is an eigenfunction of $H_1$.

Now we instead consider $\ev{A}{0}$. $\ev{A}{0}$ in the Pauli basis yields:
\begin{equation}
\begin{split}
\varepsilon^2 N_{tot}^{GPB} &= \left( \sqrt{\Var{\roothalf X}} + \sqrt{\Var{\roothalf Z} } \right)^2 \\
&= \left(  \sqrt{\half(1-\ev{X}{0}^2)} + \sqrt{\half(1-\ev{Z}{0}^2)} \right)^2 \\
&= \left(  \sqrt{\half(1-1)} + \sqrt{\half(1-0)} \right)^2  \\
&= ( 0 - \roothalf )^2 \\
&= \half
\end{split}
\end{equation}

Measuring $\ev{A}{0}$ in the RB basis $P_{X+Z}$ yields:
\begin{equation}
\begin{split}
\varepsilon^2 N_{tot}^{RB} &=  \Var{\ev{P_{X+Z}}{0}}  \\
&= ( 1 - \ev{P_{X+Z}}{0}^2 ) \\
&= ( 1 - (\roothalf)^2 ) \\
&= \half
\end{split}
\end{equation}

where we have used
\begin{equation}
P_{X+Z} = \roothalf
\begin{bmatrix}
1 & 1 \\
1 & -1
\end{bmatrix}
\end{equation}

So in the particular case of $\ev{H_1}{0}$, we see that $N_{tot}^{GPB} = N_{tot}^{RB}$. Both bases yield the same measurement counts in this case.

We have seen in the two examples above that $M_{RB}$ is either less than or equal to $M_{GPB}$. We now prove that this is the case for any one-qubit real operator and real state. The purpose of this is to show that moving to the rotated basis (RB) is always beneficial or neutral, in terms of shot counts, for this general case.

What follows is a proof of Result \ref{thm:1q}. 
\begin{proof}
Expressing a general real state as
$$
\ket{\psi_\alpha} = \alpha\ket{0} + \sqrt{1-\alpha^2}\ket{1}
$$ 
it follows that
\begin{equation}
\begin{split}
\ev{H(\eta)}{\psi_\alpha}= \alpha(\alpha\sqrt{1-\eta^2} \eta\sqrt{1-\alpha^2}) \\
+ \sqrt{1-\alpha^2}(\alpha\eta - \sqrt{1-\alpha^2}\sqrt{1-\eta^2} ) ).
\end{split}
\end{equation}
We have
$$
\varepsilon^2 M^{RB} = \ev{P_H(\eta)}{\psi_\alpha} = \ev{H(\eta)}{\psi_\alpha}
$$
where the second equality is true simply because the one-qubit Hamiltonian $H$ has unit norm. Next we write down
\begin{equation}
\begin{split}
\varepsilon^2 M^{GPB} = \left( \sqrt{\Var{\eta X} } + \sqrt{\Var{\sqrt{1 - \eta^2} Z} } \right)^2 \\
= \left( \eta \sqrt{\Var{ X} } + \sqrt{1 - \eta^2}\sqrt{\Var{Z} } \right)^2 \\
\end{split}
\end{equation}

Straightforward algebra yields
\begin{equation}
\begin{split}
\varepsilon^2 M^{GPB} - \varepsilon^2 M^{RB} \\
= (8\alpha^3\eta - 4\alpha\eta ) \sqrt{1-\alpha^2}\sqrt{1-\eta^2}  \\
+ 8\eta\sqrt{1-\eta^2}\sqrt{-\alpha^4 + \alpha^2}\sqrt{\alpha^4 - \alpha^2 + \frac{1}{4} } ,
\end{split}
\end{equation}

which is $\geq$0 for all $\eta,\alpha \in [0,1]$.

\end{proof}

\bibliographystyle{unsrt}
\bibliography{refs}

\end{document}